\documentstyle[12pt]{article}
\newcommand{\beq}{\begin{equation}}
\newcommand{\eeq}{\end{equation}}
\newcommand{\bea}{\begin{eqnarray}}
\newcommand{\eea}{\end{eqnarray}}
\newcommand{\nin}{\noindent}
\begin{document}
\begin{titlepage}
\title{Intermittency in a single event
       \thanks{e-mail: $bialas@thp1.if.uj.edu.pl$,
                       $beataz@ztc386a.if.uj.edu.pl$} }
\author{A. Bialas $^{\dag)*)}$ and B. Ziaja $^{\dag)}$}
\date{February 1996}
\maketitle
\begin{center}
$^{\dag}$ Institute of  Physics, Jagiellonian University\\
        Reymonta 4, 30-059 Cracow, Poland\\
\vspace{0.5cm}
$^{*}$ LPTHE, Universite Paris-Sud\\
        91405 Orsay Cedex, France,\\
        Laboratoire associe au CNRS URA$-$D0063\\
\end{center}
{\abstract
The possibility to study intermittency in a single event of high multiplicity is
investigated in the framework of the $\alpha-$model. It is found that, for cascade long
enough, the dispersion of intermittency exponents obtained from individual events is fairly small.
This fact opens the possibility to study the distribution of the intermittency
parameters characterizing the cascades seen (by observing intermittency)
 in particle spectra.}
\end{titlepage}


\addtolength{\baselineskip}{0.20\baselineskip}

\noindent
{\bf {1.}} The original suggestion of intermittent behaviour in multiparticle production
at high energies \cite{l1} was based on analysis of a single event of very high multiplicity
recorded by the JACEE collaboration \cite{l2}. It was soon realized, however, that
the idea can be applied to events of any multiplicity provided that a proper
averaging of the distributions is performed \cite{l3}. This led to many successful
experimental studies of intermittency \cite{l4}, and allowed to express the effect
in terms of the multiparticle correlation functions\cite{l5}. It should be realized, however, that
the averaging procedure, apart from clear advantages, brings also a danger of
overlooking some interesting effects if they are present only in a part of events produced
in high-energy collisions. It seems therefore interesting to study intermittency
parameters of individual events, hoping that they may indicate some specific production
mechanism (a typical example is the production of quark-gluon plasma which is expected to
be characterized by specific intermittency exponents, see e.g. \cite{l6},
and certainly not expected to be present in each event).

Such studies should necessarily be restricted to high-multiplicity events because only
there one may expect the statistical fluctuations to be under control. However, even
neglecting statistical errors due to the finite number of particles, there remains an
intrinsic uncertainty of the intermittency parameters: the cascade responsible for
intermitent behaviour has different realizations in different events. As the intermittency
exponents determined from different realizations of the same random cascade are expected
to scatter around the average, the method has a finite resolution with respect to the
parameters of the random cascade. Clearly, the resolution is a function of the number of
steps in the cascade.

In the present paper we investigate the distribution of intermittency exponents
obtained from analysis of individual events, using as a tool the $\alpha-$model of
one-dimensional random cascade \cite{l1}. We concentrate on two problems~:

(a) how much the average value of an intermittency exponent obtained from analysis of
individual events differs from its "theoretical" value calculated from the assumed
parameters of the random cascade and from the "standard" value obtained by averaging
factorial moments over many events.

(b) what is the dispersion of this distribution or, in other words, what is the resolution
of the measurement and how it depends on the number of steps in the cascade.

\noindent
We have investigated cascades of up to 10 steps. Our results are described in some detail
below. They can be summarized as follows~:

(i) the average value of the intermittency exponent reproduces well the value obtained by
averaging the factorial moments over events. This result weakly depends on
the number of steps in the cascade. However, both numbers have a tendency to underestimate
the theoretical value for cascades of short lengths.

(ii) the dispersion of the distribution is inversely proportional to the length of the
cascade. For cascades of less than 6 steps it becomes fairly large, but still substantially smaller
than the average value.

We thus find that a measurement of the distribution of intermittency exponents
obtained from individual events of high multiplicity may indeed help to reveal existence
of groups of events emerging from cascades with different characteristics.
\\
\\
\noindent
{\bf {2.}} The $\alpha-$model of random cascading \cite{l1} describes a multiparticle event
as a series of steps in which each phase-space interval is divided into some number
of equal
parts. At any step n ( $n=1,2,\ldots,N$) particle density in each of the parts
is
obtained by multiplication of the density at the (n-1)th step by one of the two values
(a,b) of random variable W with the probabilities $\alpha$ and $\beta$,
respectively.
For simplicity one assumes also~:
\beq
<W>=\alpha a + \beta b =1
\eeq
\noindent
where $<>$ denotes the average value henceforth. Note that (1) implies~:
\beq
\alpha=\frac {b-1}{b-a},\, \beta =\frac {1-a}{b-a}
\eeq
So that the model is defined by two parameters $a$ and $b$.

In our simulation of the $\alpha$-model we have divided each bin into
2 parts,
so the number of bins at each step equals~:
\beq
M(n) = 2^{n}
\eeq
and thus the length of each bin is equal to~:
\beq
d(n)=\frac{D}{M(n)}
\eeq
\noindent
where D is the total phase-space interval.

The "standard" method is to study scaling behaviour of the normalized moments of particle
densities~:
\beq
< Z^{q}_{m}(d)>=< (x_{m}(d))^{q} >
\eeq

\nin
Here $x_{m}(d)$ is the density obtained after n steps of the cascade in the mth
bin
( $m=1,\ldots,M(n)$ ), and the average is taken over all considered events. It follows
from (1) that $< Z^{1}_{m} >=1 $. In the $\alpha-$ model $< Z^{q}_{m}(d)>$
follows the power law~:
\beq
<Z^{q}_{m}(d)>=(\frac{2D}{d})^{\varphi_{q}}
\eeq
\nin
where the intermittency exponents are given by~:
\beq
\varphi_{q}=\log_{2} <W^{q}>
\eeq
\\
\noindent
{\bf {3.}} If one is interested in event-by-event analysis, one is forced to consider the so-called
horizontal average $Z^{q}(d)$ \cite{l1}~:
\beq
Z^{q}(d)=M^{-1} \sum_{m=1}^{M} Z_{m}^{q}(d)
\eeq
obtained by averaging over all bins. In the $\alpha-$model the average particle density is
independent of m and thus $Z^{q}(d)$ follows the same scaling law (6)
as $Z^{q}_{m}(d)$~:
\beq
Z^{q}(d)=(\frac{2D}{d})^{\varphi_{q}}
\eeq

As we have already explained, $\varphi_{q}$ calculated from (9) fluctuate from event
to event even for fixed parameters of the cascade. Its average over many events
should approach the value given by (7). The dispersion around the average,
however, does not vanish, even in the limit of infinite number of events.
In other words, even for events with very large multiplicity we cannot determine
intermittency exponents with arbitrarily high precision: there is a "natural"
uncertainty of this measurement. This uncertainty is expected to decrease with
increasing number of steps in the cascade. Furthermore, the dispersion of
the distribution of the factorial moment $Z_{q}(d)$ can be estimated as:
\beq
D^{2}(Z^{q}(d)) \simeq const
\eeq
which explicitly shows that $D(\varphi_{q})$ is inversely proportional to the
length of the cascade.

We have performed numerical simulations of the $\alpha$-model in order to obtain
the feeling to what extent these theoretical prejudices are realized in practice.
In Figs.\ 1, 2 the histograms of the values of  intermittency exponent $\varphi_{2},\varphi_{3}$
are plotted
for  5000 generated cascades with 6 and 10 steps. These numbers were determined
for each event using the method described in \cite{l1}
(and applied there to the JACEE event \cite{l2}). One sees that both the average
value and the dispersion depend on number of steps in the cascade.
For small number of steps the average value obtained from simulation is
smaller than the "true" value given by Eq.\ (7). For 10 steps,
however, the simulation gives the average rather close to the theoretical
result.
The dispersion of the distribution estimated directly from the observed peak,
decreases with the number of cascade steps following well the $1/n$
rule of the Eq.(10). Its numerical value as a function of the cascade length
is presented in Tables 1, 2 for 2 different sets of cascade parameters a, b.
The dispersion is relatively small, and it allows
to distinguish between the cascades with different parameters (Figs.1, 2).\\

\begin{figure}[hbpt]
\noindent
Table 1.\ Intermittency exponents and their dispersions for $a=0.8,\, b=1.1$
and $n=5,\ldots,10$ cascade steps

\begin{center}
\begin{tabular}{|l|c|c|c|c|c|c|c|}
\hline \hline
                           & theor.& 5         &    6       &      7    &    8      &    9       &   10 \\
\hline
$\varphi_{2}=10^{-2}\times$& 2.9  &$2.4\pm0.9$&$2.5\pm 0.8$&$2.6\pm0.7$&$2.7\pm0.6$&$2.7\pm 0.6$&$2.7\pm0.5$\\
\hline
$\varphi_{3}=10^{-2}\times$& 8.2  &$6.9\pm2.6$ &$7.1\pm 2.3$&$6.6\pm2.0$&$7.7\pm1.7$&$7.7\pm 1.6$&$7.8\pm1.5$\\
\hline \hline
\end{tabular}
\end{center}
\end{figure}
\begin{figure}[hbpt]
\noindent
Table 2.\ Intermittency exponents and their dispersions for $a=0.5,\, b=1.5$
and $n=5,\ldots,10$ cascade steps
\begin{center}
\begin{tabular}{|l|c|c|c|c|c|c|c|}
\hline \hline
                           & theor.&5          &    6      &    7      &    8      &    9      &   10   \\
\hline
$\varphi_{2}=10^{-1}\times$& 3.2  &$2.4\pm1.0$&$2.5\pm1.0$&$2.5\pm0.8$&$2.7\pm0.7$&$2.7\pm0.6$&$2.8\pm0.6$\\
\hline
$\varphi_{3}=10^{-1}\times$& 8.1  &$5.9\pm2.3$&$6.1\pm2.2$&$6.4\pm2.1$&$6.7\pm1.8$&$6.7\pm1.6$&$6.8\pm1.6$\\
\hline \hline
\end{tabular}
\end{center}
\end{figure}

\noindent
{\bf {4.}} Our conclusions can be summarized as follows~:\\

(a) the average value of the intermittency exponent obtained from our analysis
is fairly close to the "theoretical" value. \\

(b) the dispersion of the distribution is inversely proportional to the
length of the cascade. It is found to be relatively small. This allows
to distinguish between cascades with reasonably different parameters.\\

\noindent
{\bf Acknowledgements}\\
The research was supported in part by the KBN grant No. 2 PO 3B 08308 and
by the European Human Capital and Mobility Program ERBCIPDCT940613.

\newpage

\noindent
\underline {Figure captions}\\

\noindent
{\bf Fig.\ 1 } Distribution of the intermittency exponent $\varphi_{2}$
as determined in individual  events generated from the $\alpha-$model.
5000 events with $a=0.8$, $b=1.1$ (a) and 5000 events with $a=0.5$,
$b=1.5$ (b) were used. Cases (a) and (b) are plotted in two different
scales. Histogram for the case (a) is multiplied by $10^{-1}$.\\
Solid line and dots~: 10 cascade steps, dashed line and crosses~:
6 cascade steps.\\

\noindent
{\bf Fig.\ 2 } Distribution of the intermittency exponent $\varphi_{3}$.
Other details as in Fig.\ 1.

\end{document}